\documentclass[aip,pop,preprint,superscriptaddress]{revtex4-1}
\usepackage{setspace}
\usepackage{amsfonts,amsmath,amssymb}
\usepackage{txfonts}
\usepackage{graphics}
\usepackage{bm}
\usepackage{verbatim}
\usepackage{hyperref}
\usepackage{units}
\usepackage[T1]{fontenc}
\usepackage[latin9]{inputenc}
\usepackage{amsbsy}
\usepackage{esint}

\begin{document}

\title{Gauge properties of the guiding center variational symplectic integrator}

\author{J.~Squire}
\affiliation{Plasma Physics Laboratory, Princeton University, Princeton, New Jersey 08543, USA}
\author{H.~Qin}
\affiliation{Plasma Physics Laboratory, Princeton University, Princeton, New Jersey 08543, USA}
\affiliation{Dept.~of Modern Physics, University of Science and Technology of China, Hefei, Anhui 230026, China}
\author{W.~M.~Tang}
\affiliation{Plasma Physics Laboratory, Princeton University, Princeton, New Jersey 08543, USA}

\begin{abstract}
Variational symplectic algorithms have recently been developed for carrying out long-time simulation of charged 
particles in magnetic fields\cite{Qin:2008p5812,Qin:2009p5777,Li:2011p7068}. As a 
direct consequence of their derivation from a discrete variational principle, these algorithms have 
very good long-time energy conservation, as well as exactly preserving discrete momenta. We present 
stability results for these algorithms, focusing on understanding how explicit variational integrators can 
be designed for this type of system. It is found that for explicit algorithms an instability arises because the discrete symplectic 
structure does not become the continuous structure in the $t \rightarrow 0$ limit.
We examine how a generalized gauge transformation can be used to put the Lagrangian in 
the "antisymmetric discretization gauge," in which the discrete symplectic structure 
has the correct form, thus eliminating the numerical instability. Finally, it is noted that the variational guiding 
center algorithms are not electromagnetically gauge invariant. By designing a model 
discrete Lagrangian, we show that the algorithms are approximately gauge invariant as long as $\bm{A}$  and $\phi$ are relatively smooth. 
A gauge invariant discrete Lagrangian is very important in a variational particle-in-cell algorithm  
where it ensures current continuity and preservation of Gauss's law\cite{PRLtobePublished}.
\end{abstract}


\date{\today}

\maketitle

In many applications involving magnetized plasmas, it is necessary
to numerically integrate particle dynamics over long time scales.
A crucial associated tool is the guiding center description \cite{Littlejohn1983},
which averages over fast gryomotion, allowing a dramatic decrease
in necessary computational resources through the use of much longer
time steps \cite{Lin18091998,Chen2003463,Cohen198045}. Traditional integration routines (for
instance Runga-Kutta) for the guiding center equations, while much
more efficient than integration of the full Lorentz force equations,
can perform badly over very long simulation times. To mitigate these
problems and improve confidence in simulation results, variational
integrators for the guiding center equations have recently been presented
in Refs.~\onlinecite{Qin:2008p5812,Qin:2009p5777,Li:2011p7068}. Based on a discretization
of the variational principle rather than the equations of motion \cite{springerlink:10.1007/BF01077598},
these algorithms exactly conserve a symplectic structure \cite{0951-7715-3-2-001,Candy1991230,Feng:1986:DSH:8528.8538,Yoshida:1990p5808}.
As a consequence\cite{Marsden:2001varint,0951-7715-3-2-001,Reich:1999:BEA:333858.333895,springerlink:10.1007/BF01077598} 
they exhibit very good long time conservation properties, and numerical
solutions stay close to exact dynamics, even at large time-step. In
addition, a discrete Noether's theorem implies that exact numerical
conservation laws arise from symmetries of the system, for instance
momentum conservation due to translational invariance. 

The basic idea behind variational integrators is simple and represents a departure from 
the usual approach of 
deriving continuous equations of motion from a continuous Lagrangian and 
then discretizing these differential equations. Instead, the Lagrangian itself is discretized and
an integrator is derived from this \emph{discrete} variational principle
\cite{Marsden:2001varint}. In this process, there is of course some freedom
in the chosen discretization of the Lagrangian. For example, $\bm{x}\left(t\right)$  could 
be discretized as $\bm{x}_k$, or as $\frac{1}{2}\left(\bm{x}_k+\bm{x}_{k+1}\right)$.
In this paper we investigate a different type of freedom that has previously not
been studied (to our knowledge) -- \emph{the freedom to gauge transform 
the continuous Lagrangian}.
It is well known that a generalized
gauge transformation, $L\rightarrow L+\frac{d}{dt}S$, does not change
the continuous Euler-Lagrange equations of motion. Nevertheless, 
the discrete Euler-Lagrange equations
derived from a discretization of $L$ are in general not the same
as those from a discretization of $L+\frac{d}{dt}S$. This article 
presents the results of a systematic investigation of 
the effects of these gauge transformations on the properties
of the variational symplectic guiding center algorithms. In particular,
we find that the choice of gauge can profoundly alter the algorithms'
stability properties. These results are intended to be a guide for
future users of the guiding center algorithms, as well as variational
integrators for systems with Lagrangians of a similar form -- such as
the magnetic field line Lagrangian \cite{Cary_Littlejohn_1983} or
point vortices \cite{Rowley2002}.

The lowest order non-canonical Lagrangian for the guiding center system,
given by Grebogi and Littlejohn \cite{grebogi:1996,Littlejohn1983},
is \begin{equation}
L=\left[\bm{A}\left(\bm{x}\right)+U\bm{b}\left(\bm{x}\right)\right] \cdot 
\dot{\bm{x}}+\mu\dot{\Theta}-\left[\phi\left(\bm{x}\right)+\Gamma\left(\bm{x},U\right)\right].\label{eq:Full GC Lagrangian}
\end{equation}
Here $\bm{x}$ is the guiding center position, $U=\gamma u$ is
the relativistic  momentum parallel to the magnetic field (with $\gamma$ the relativistic
mass factor), $\mu$ is the conserved magnetic moment, $\Theta$ is
the gyrophase, $\bm{b}\left(\bm{x}\right)$ is the magnetic
field unit vector, $\bm{A}\left(\bm{x}\right)$ is the magnetic
vector potential, $\phi\left(\bm{x}\right)$ is the electric potential
and $\Gamma\left(\bm{x},U\right)=\sqrt{1+U^{2}+2\mu B\left(\bm{x}\right)}.$
Note $\bm{A}$ is normalized by $c/e$, $\Gamma$ by $1/mc^{2}$
and $\phi$ by $1/e$. In the non-relativistic limit, $U$ becomes
$u$ (parallel velocity) and $\Gamma$ becomes $1+u^{2}/2+\mu B\left(\bm{x}\right)$.
Since only the time derivative of the gyrophase ($\Theta$) appears
in Eq.~\eqref{eq:Full GC Lagrangian}, the equation of motion for
$\mu$ is simply $\mu \left(t\right) = \mu \left(0\right)$ and we ignore this term in
the Lagrangian for the remainer of the article. Continuous particle guiding center
equations of motion are derived as usual from Eq.~\eqref{eq:Full GC Lagrangian} with the Euler-Lagrange equations.

The variational symplectic guiding center algorithms in Refs.~\onlinecite{Li:2011p7068,Qin:2008p5812,Qin:2009p5777}
are derived from discretizations of Eq.~\eqref{eq:Full GC Lagrangian}.
We give a brief overview of this process for clarity.  For the algorithm of 
Refs.~\onlinecite{Qin:2008p5812,Qin:2009p5777} the (non-relativistic) discrete Lagrangian is chosen
to be,
\begin{align}
L_{d}\left(k,k+1\right)= & \frac{1}{2}\left[\bm{A}^{\dagger}\left(\bm{x}_{k}\right)+\bm{A}^{\dagger}\left(\bm{x}_{k+1}\right)\right]\cdot\frac{\left(\bm{x}_{k+1}-\bm{x}_{k}\right)}{h}\nonumber \\
- & \left[\frac{u_{k}u_{k+1}}{2}+\mu B\left(\bm{x}_{k}\right)+\phi\left(\bm{x}_{k}\right)\right],\label{eq:Discrete L example}\end{align}
where $\bm{A}^{\dagger}\left(\bm{x}\right)\equiv \bm{A}\left(\bm{x}\right)+u\,\bm{b}\left(\bm{x}\right)$.
Eq.~\eqref{eq:Discrete L example} is a direct approximation of $\frac{1}{h}\int_{t_{k}}^{t_{k+1}}dt\, L\left(\bm{x},\dot{\bm{x}},U,\dot{U}\right)$.
Requiring stationarity of the discrete action $\mathcal{A}_{d}=\sum_{k}hL_{d}$
under arbitrary variations $\left(\delta\bm{x}_{k},\,\delta u_{k}\right)$
$\left(0<k<N\right)$, leads to the discrete update equations for the
system,
 \begin{align}
\frac{1}{2h}&A_{i,j}^{\dagger} \left(\bm{x}_{k}\right)\left(x_{k+1}^{i}-x_{k-1}^{i}\right)-\frac{1}{2h}\left[A_{j}^{\dagger}\left(\bm{x}_{k+1}\right)-A_{j}^{\dagger}\left(\bm{x}_{k-1}\right)\right]\nonumber \\
&= \, \mu B_{,j}\left(\bm{x}_{k}\right)+\phi_{,j}\left(\bm{x}_{k}\right)\qquad\left(j=1,\,2,\,3\right),\label{eq:VSI algorithm 1} \\
\frac{1}{2h}&b_{i} \left(\bm{x}_{k}\right)\left(x_{k+1}^{i}-x_{k-1}^{i}\right)=\frac{1}{2}\left(u_{k+1}+u_{k-1}\right).\label{eq:VSI algorithm 2}
\end{align}
These equations are solved implicitly to integrate particle motion
through phase space. For the purposes of this article, the discretization of Eq.~\eqref{eq:Discrete L example},
$\bm{A}^{\dagger}\left(\bm{x}\right)\rightarrow\frac{1}{2}\left[\bm{A}^{\dagger}\left(\bm{x}_{k}\right)+\bm{A}^{\dagger}\left(\bm{x}_{k+1}\right)\right]$
is equivalent to $\bm{A}^{\dagger}\left(\bm{x}\right)\rightarrow\bm{A}^{\dagger}\left(\frac{1}{2}\left(\bm{x}_{k}+\bm{x}_{k+1}\right)\right)$
(used in Ref.~\onlinecite{Li:2011p7068}) since our analysis is carried
out on the linearized system. 

This paper presents results on the stability of the variational symplectic guiding center 
algorithms. We carry out analysis to determine whether an explicit variational integrator 
can be designed. It is found that in general, explicit integrators are numerically unstable 
at any time step. This instability is shown to be a direct result of the relationship between 
the conserved symplectic structure of the continuous Euler-Lagrange equations and that 
of the discrete integrator. The reduction of the symplectic 2-form basis from 
$\mathbf{d}x^{\nu}_{k} \wedge\mathbf{d}x^{\mu}_{k+1}$ to $\mathbf{d}x^{\nu}\wedge\mathbf{d}x^{\mu}$ 
in the limit of zero time-step can lead to differences between the discrete and continuous structures, causing an instability. 
This knowledge leads to a way to eliminate the instability in some cases, by using 
a generalized gauge transformation of the Lagrangian to the "antisymmetric discretization gauge". 
Such an approach ensures that the discrete 
symplectic structure becomes the continuous structure as $t\rightarrow 0$. 
The idea that  gauge transformations can profoundly alter 
stability properties of variational algorithms leads to an important realization that 
merits further investigation. Due to the discretization schemes adopted, the variational symplectic 
guiding center integrators reported in Refs.~\onlinecite{Qin:2008p5812,Qin:2009p5777,Li:2011p7068} are 
not electromagnetically gauge invariant, even though the continuous Lagrangian is gauge invariant. 
This implies that integrated particle dynamics depend 
on the details of $\bm{A}$ and $\phi$, not just $\bm{B}=\nabla \times \bm{A}$ and 
$\bm{E} = -\partial_t \bm{A}-\nabla \phi$. We examine the importance of this by first designing 
a gauge invariant variational integrator and comparing this to the algorithms in 
Refs.~\onlinecite{Qin:2008p5812,Qin:2009p5777,Li:2011p7068}. This method illustrates that 
as long as $\bm{A}$ and $\phi$ are relatively smooth (in comparison to a particle step), the 
algorithm is approximately electromagnetically gauge invariant and integrated particle dynamics 
should be accurate. These ideas are important for the design of variational particle-in-cell 
schemes, since a gauge invariant discrete Lagrangian ensures current continuity and exact 
preservation of Gauss's law\cite{PRLtobePublished}.

In Section~\ref{sec:Discretization-Gauge-and}
we outline the symplectic properties of the guiding center variational integrators 
and examine linear stability. These ideas are used to design the 
antisymmetric discretization gauge, in which explicit integrators are stable.
Electromagnetic gauge transformations are examined
in Section~\ref{sec:Electromagnetic-gauge}, where it is demonstrated that smooth $\bm{A}$ and $\phi$ 
ensure approximate gauge invariance and accurate integration of particle trajectories.
Illustrative numerical examples are given in both sections.

\section{Discretization Gauge and linear stability\label{sec:Discretization-Gauge-and}}

In this section it is most instructive to consider a generic non-canonical
Lagrangian of the form,\begin{equation}
L\left(q,\dot{q}\right)=\left\langle \gamma\left(q\right),\dot{q}\right\rangle -H\left(q\right).\label{eq:Generic degenerate L}\end{equation}
Here $\gamma\left(q\right)$ is a 1-form and $H\left(q\right)$ is
a function, both on the phase space $Q$. The guiding center Lagrangian,
Eq.~\eqref{eq:Full GC Lagrangian}, is of this form, with $q=\left(\bm{x},\, U\right)$,
$\gamma=\left[A_{i}\left(\bm{x}\right)+Ub_{i}\left(\bm{x}\right)\right]dx_{j}$,
$j=\left(1,\,2,\,3\right)$ and $H=\phi+\Gamma$. Properties of variational
integrators for Lagrangians of this form have also been studied in
the context of vortex dynamics in Ref.~\onlinecite{Rowley2002}.

\subsection{Symplectic structure}

To better understand the characteristics of the variational guiding center
algorithm, we first discus some curious attributes of the Lagrangian
Eq.~\eqref{eq:Generic degenerate L}. The usual conserved symplectic 
structure is defined on the tangent bundle of the phase space, $TQ$, and is 
given in co-ordinates by\cite{Marsden:2001varint} 
\begin{equation}
\Omega_{L}=\frac{\partial^{2}L}{\partial\dot{q}^{i}\partial q^{j}}\mathbf{d}q^{i}\wedge\mathbf{d}q^{j}+\frac{\partial^{2}L}{\partial\dot{q}^{i}\partial\dot{q}^{j}}\mathbf{d}q^{i}\wedge\mathbf{d}\dot{q}^{j}.\label{eq:Usual Lag sym structure}
\end{equation}
This is degenerate if the matrix $\partial^{2}L/\partial\dot{q}^{i}\partial\dot{q}^{j}$
is singular, which is the situation for Lagrangians of the form of Eq.~\eqref{eq:Generic degenerate L}.
In this case it makes little sense to describe the Euler-Lagrange flow as
being symplectic on $TQ$, since by definition a symplectic structure
is non-degenerate. However, for the particular form of the Lagrangian
in Eq.~\eqref{eq:Generic degenerate L} there is a conserved structure
on the phase space, $Q$, which will turn out to be very important for the stability
of the discretization. The existence of such a structure is shown by considering
the action integral $\mathcal{S}=\int_{0}^{t}L\left[q\left(t'\right),\dot{q}\left(t'\right)\right]dt'$. For $q\left(t\right)$ that satisfies the
Euler-Lagrange equations, taking the exterior derivative leads to\cite{Rowley2002} 
\begin{equation}
\mathbf{d}\mathcal{S}=\left.\frac{\partial L}{\partial\dot{q}^{i}}dq^{i}\right|_{0}^{t}=\left.\gamma_{i}dq^{i}\right|_{0}^{t}=F_{t}^{*}\gamma-\gamma,\label{eq:Sym form derivation}\end{equation}
where $F_{t}^{*}$ is the flow map. Using $\mathbf{d}^{2}=0$ gives
\begin{equation}
F_{t}^{*}\mathbf{d}\gamma=\mathbf{d}\gamma\label{eq:sym form conservation}\end{equation}
showing that $-\mathbf{d}\gamma$ is a symplectic structure (on $Q$
rather than $TQ$) conserved by the flow of the Euler-Lagrange equations.
Note that for this type of degeneracy, the Euler-Lagrange
equations are first order in time. 

We now consider discretizations of Eq.~\eqref{eq:Generic degenerate L},
in which case we have discrete equations of motion on $Q\times Q$.
For concreteness, all discretizations used in this section simply
replace $q\left(t\right)$ with \begin{equation}
q_{\alpha}=\left(1-\alpha\right)q_{k}+\alpha q_{k+1},\label{eq:Discretization defn}\end{equation}
with $0\leq\alpha\leq1$, and $\dot{q}\left(t\right)$ with $\left(q_{k+1}-q_{k}\right)/h$
to create a discrete Lagrangian ($h$ denotes the time-step). This
is identical to the variational guiding center algorithm in Ref.~\onlinecite{Li:2011p7068}
and very similar to that in Refs.~\onlinecite{Qin:2008p5812,Qin:2009p5777}, with
results holding for both of these cases since our analysis is linear. 
For a discrete Lagrangian $L_{d}\left(q_{k},q_{k+1}\right)$,
the discrete Euler-Lagrange equations, derived by requiring 
stationarity of the action under arbitrary variations, $\delta q_k$, are given by\begin{equation}
\frac{\partial}{\partial q_{k}}\left[L_{d}\left(q_{k-1},q_{k}\right)+L_{d}\left(q_{k},q_{k+1}\right)\right]=0.\label{eq:Discrete EL equations}\end{equation}
The discrete symplectic structure,\begin{equation}
\Omega_{L_{d}}=\frac{\partial^{2}L_{d}}{\partial q_{k}^{i}\partial q_{k+1}^{j}}\mathbf{d}q_{k}^{i}\wedge\mathbf{d}q_{k+1}^{j},\label{eq:Discrete sym form}\end{equation}
is preserved by the flow of the \emph{discrete} Euler-Lagrange map; i.e., the discrete 
update equations for the integrator.
Degeneracy of the continuous Lagrangian on $TQ$ (i.e., degeneracy
of $\Omega_{L}$ [Eq.~\eqref{eq:Usual Lag sym structure}]), \emph{does
not} imply $\Omega_{L_{d}}$ is degenerate on $Q\times Q$. For all
cases examined in this article $\Omega_{L_{d}}$ is non-degenerate. The stability
results presented are related to how $\Omega_{L_{d}}$ becomes the
symplectic form on $Q$ (ie.~$-\mathbf{d}\gamma$) in the $h\rightarrow0$
limit.

\subsection{Linear stability}

The variational guiding center algorithms in Refs.~\onlinecite{Qin:2009p5777,Qin:2008p5812,Li:2011p7068} 
use a discretization of $\gamma$
that is symmetric in $q_{k}$ and $q_{k+1}$ (this corresponds to
$\alpha=1/2$ in Eq.~\eqref{eq:Discretization defn}). As a consequence, 
the update equations are implicit in $q_{k+1}$, and the question
naturally arises as to whether an \emph{explicit} variational integrator
can be designed. We examine this issue by studying the stability of the
discretization of Eq.~\eqref{eq:Generic degenerate L} as the parameter
$\alpha$ [Eq.~\eqref{eq:Discretization defn}] is varied. An algorithm
is explicit for $\alpha=0$. The standard technique for numerical
stability analysis of nonlinear integrators is to calculate stability
boundaries with $\dot{x}=\lambda_{i}x$ for the algorithm in question,
where $\lambda_{i}$ are the eigenvalues of the Jacobian matrix at
some point. This technique does not carry over easily to variational
integrators, since the algorithm is defined by the discrete Lagrangian,
and accordingly cannot be easily applied to $\dot{x}=\lambda_{i}x$. Instead, we
consider a general linearization of the discrete equations of motion,
which can be represented by the equations of motion arising from 
a discrete Lagrangian of the form,\begin{equation}
L_{d,lin}=\frac{1}{h}\left(x_{k+1}^{\mu}-x_{k}^{\mu}\right)G_{\mu\nu}\, x_{\alpha}^{\nu}-x_{\alpha}^{\mu}B_{\mu\nu}\, x_{\alpha}^{\nu}-B_{L,\mu}x_{\alpha}^{\mu},\label{eq:Linearized discrete lagrangian}\end{equation}
where the summation convention is used and greek indices run $1\rightarrow4$
(including $U$). The constant matrices $G_{\mu\nu}$, $B$ and $B_{L}$
could be calculated explicitly for specific forms of $\bm{A}\left(\bm{x}\right)$
and $\phi\left(\bm{x}\right)$ (at some point) if desired. Here we consider
them to be general, with the last row of $G_{\nu\mu}$ equal to zero
(since this is the form of the guiding center Lagrangian). Note
that $B$ and $B_{L}$ contain quadratic approximations to both $\phi\left(\bm{x}\right)$
and $\Gamma\left(\bm{x},U\right)$, but these turn out to be unimportant. 
The general equations of motion arising from such a Lagrangian are in 
the form of a linearization of a discretization of Eq.~\eqref{eq:Full GC Lagrangian} about any point.
Consequently, we consider stability of the algorithm obtained
from Eq.~\eqref{eq:Linearized discrete lagrangian} to be a necessary
condition for stability of the variational guiding center integrator.
With the discrete Euler-Lagrange equations Eq.~\eqref{eq:Discrete EL equations},
we can derive the equations of motion for the linearized system in
the form \[
x_{k+1}^{\nu}=P_{\nu\mu}\left(\alpha\right)x_{k}^{\mu}+Q_{\nu\mu}\left(\alpha\right)x_{k-1}^{\mu},\]
where $P$ and $Q$ are constant matrices with dependence on $\alpha$,
$G_{\mu\nu}$, $B_{\mu\nu}$ and $B_{L,\mu}$. Stability properties
follow from the eigenvalues of this equation, given by \begin{equation}
\det\left[\lambda_{i}^{2}I-\lambda_{i}P-Q\right]=0.\label{eq:linear e-value eqn}\end{equation}
Calculating these eigenvalues in the limit $h\rightarrow0$ for arbitrary
$G_{\mu\nu}$ ($B_{\mu\nu}$ and $B_{L,\mu}$ do not contribute in
this limit), leads to $\lambda_{i}=1$, a series of $\lambda_{i}$
that depend on $G_{\mu\nu}$, and \begin{equation}
\lambda_{i}=\frac{1-\alpha}{\alpha},\:\frac{\alpha}{1-\alpha}.\label{eq:e-values for algorithm}\end{equation}
These final two eigenvalues indicate that the algorithm will be unstable
unless $\mathrm{Re}\left(\alpha\right)=\nicefrac{1}{2}$, demonstrating
an explicit scheme ($\alpha=0$) is unstable at all timesteps. In order to: (i) 
understand the reason for this behaviour, and (ii) design explicit integrators
under certain conditions, we consider gauge transformations and the
symplectic form.

\subsection{The discretization gauge}

The $h\rightarrow0$ limit of $\Omega_{L_{d}}$ [Eq.~\eqref{eq:Discrete sym form}]
for the general discrete Lagrangian [Eq.~\eqref{eq:Generic degenerate L}]
is \cite{Rowley2002},\begin{equation}
\Omega_{L_{d}}\left(x_{k},x_{k+1}\right)\approx\left.\left(\frac{\partial\gamma_{\mu}}{\partial x^{\nu}}-\alpha\left(\frac{\partial\gamma_{\nu}}{\partial x^{\mu}}+\frac{\partial\gamma_{\mu}}{\partial x^{\nu}}\right)\right)\right|_{x_{\alpha}}\mathbf{d}x_{k}^{\nu}\wedge\mathbf{d}x_{k+1}^{\mu}.\label{eq:h 0 limit, discrete sym form}\end{equation}
At exactly $h=0$, $x_{k}$, $x_{k+1}$ and $x_{\alpha}$ all become
$x$ and the 2-form basis is reduced to $\mathbf{d}x^{\nu}\wedge\mathbf{d}x^{\mu}$.
Comparing this to the continuous symplectic form on $Q$, \begin{equation}
\mathbf{d}\gamma=\frac{1}{2}\left(\frac{\partial\gamma_{\mu}}{\partial x^{\nu}}-\frac{\partial\gamma_{\nu}}{\partial x^{\mu}}\right)\mathbf{d}x^{\nu}\wedge\mathbf{d}x^{\mu},\label{eq:d gamma in coords}\end{equation}
it is clear that the two expressions co-incide at $h=0$ only if $\partial\gamma_{\mu}/\partial x^{\nu}$
is antisymmetric, or if $\alpha=1/2$. Thus, the numerical
instability away from $\alpha=1/2$ can be thought of
as a direct consequence of $\Omega_{L_{d}}$ not transforming into
the continuous preserved symplectic form, $\mathbf{d}\gamma$, in
the $h\rightarrow0$ limit.

This realization also provides a method for designing integrators that work
away from $\alpha=1/2$, since if $\partial\gamma_{\mu}/\partial x^{\nu}$
is antisymmetric, we would expect the algorithm to be stable for all $\alpha$ (at $h\rightarrow0$).
Note that $\partial\gamma_{\mu}/\partial x^{\nu}$ will not be antisymmetric
for the variational guiding center algorithms; however, we can use
the fact that the continuous Euler-Lagrange equations are unchanged
by the addition of a total time derivative to the Lagrangian, a generalized gauge transformation. For
some arbitrary function $S$, this is equivalent to
$\gamma_{\mu}\rightarrow\gamma'_{\mu}=\gamma_{\mu}+S_{,\mu}$, $H\rightarrow H'=H-\partial_{t}S$
in Eq.~\eqref{eq:Generic degenerate L}. An integrator derived from
this transformed Lagrangian should simulate the same continuous dynamics,
though the discrete update equations are different. We can require $\partial \gamma'_{\mu} /\partial x^{\nu}$
be antisymmetric, which leads to the partial differential equation,\begin{equation}
S_{,\,\mu\nu}=-\frac{1}{2}\left(\gamma_{\mu,\nu}+\gamma_{\nu,\mu}\right),\label{eq:Gauge transf condition}\end{equation}
that can easily be solved for the linearized Lagrangian, Eq.~\eqref{eq:Linearized discrete lagrangian}.
Numerical tests show the algorithm to be stable for all $\alpha$ when this
antisymmetric discretization gauge ($\partial \gamma_{\mu} /\partial x^{\nu}$ 
antisymmetric) is used. Note that Eq.~\eqref{eq:Gauge transf condition} does not
always have a solution: equality of mixed third derivatives of $S$
leads to the condition \begin{equation}
\gamma_{\mu,\nu\lambda}-\gamma_{\nu,\mu\lambda}=0,\label{eq:gamma condition}\end{equation}
which is trivially satisfied for the linear case, but in general not
true globally for the guiding center Lagrangian, Eq.~\eqref{eq:Full GC Lagrangian}.
Thus, while the Lagrangian can locally be put into the antisymmetric discretization
gauge by linearizing about some point, the global gauge
may not exist for arbitrary $\gamma$. Note that if Eq.~\eqref{eq:gamma condition} 
is not satisfied, a global gauge could still exist in a different co-ordinate system. A trivial
example of this would be if canonical co-ordinates existed for the
guiding center Lagrangian of the field in question\cite{white:573},
in which case $\gamma_{j}=P_{j},\:\gamma_{j+3}=0,\; j=\left(1,\,2,\,3\right)$
and Eq.~\eqref{eq:gamma condition} is satisfied. Canonical
co-ordinates do not always exist, and it is not yet clear if
there is a co-ordinate change that would allow a global antisymmetric discretization
gauge for a general magnetic field. This interesting theoretical question will 
be investigated further in the future. For practical purposes, it is always 
possible to pick an antisymmetric discretization gauge in the neighborhood of some point.

\subsection{Numerical example}

\begin{figure}

\begin{centering}

\includegraphics{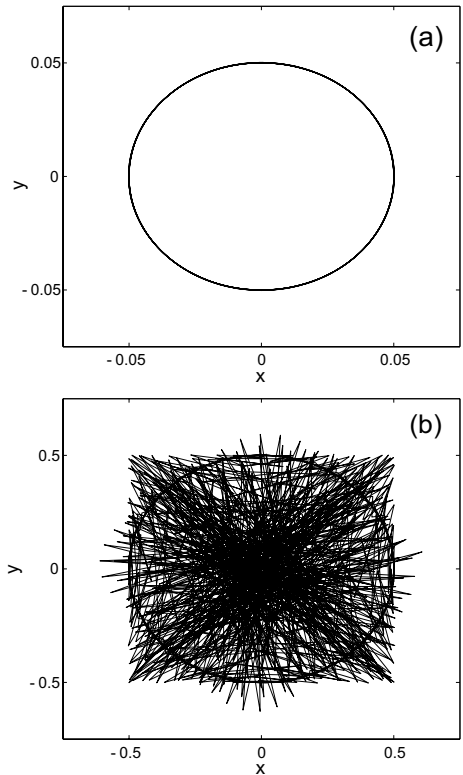}
\caption{Particle trajectories integrated explicitly ($\alpha=0$) for the
field $\bm{B}\left(\bm{x}\right)=\left[1+\left(x^{2}+y^{2}\right)/20\right]\bm{\hat{z}}$
put into the linear antisymmetric discretization gauge of the point $\left(0,0\right)$.
A trajectory that remains close to $\left(0,0\right)$ is numerically
stable (a), while further away (b), it is unstable because the local
antisymmetric discretization gauge at $\left(0,0\right)$ is not a good approximation
to the required gauge at the particle position. \label{fig:D gauge example}}
\end{centering}

\end{figure}

We now give a simple numerical example to illustrate the effect of
a transformation into a local antisymmetric discretization gauge. We use the non-relativistic
guiding center algorithm, with magnetic field\begin{equation}
\bm{B}\left(\bm{x}\right)=\left[1+\left(x^{2}+y^{2}\right)/20\right]\bm{\hat{z}},\label{eq:Num examp field}\end{equation}
in which particles execute closed circular orbits, $x^{2}+y^{2}=\mathrm{const}.$
This field can be represented by $\bm{A}^{\dagger}\left(\bm{x}\right)\equiv\bm{A}\left(\bm{x}\right)+u\bm{b\left(\bm{x}\right)}=\left(-\frac{1}{60}y^{3},\, x+\frac{1}{60}x^{3},\, u,\,0\right)$,
including the $u$ component (since this is needed when we change
gauges). There is no global antisymmetric discretization gauge for this field, as
Eq.~\eqref{eq:gamma condition} cannot be globally satisfied. However, since
particles orbit around $\left(x,y\right)=\left(0,0\right)$, we can
choose the local gauge associated with linearization of the
equations of motion around this point. This corresponds to $S=-\frac{1}{2}xy-\frac{1}{2}zu$,
giving $\bm{A}^{\dagger}\left(\bm{x}\right)$ in the new gauge
as, \begin{equation}
\bm{A}'^{\dagger}\left(\bm{x}\right)=\left(-\frac{1}{60}y^{3}-\frac{y}{2},\,\frac{1}{60}x^{3}+\frac{x}{2},\,\frac{u}{2},\,-\frac{z}{2}\right).\label{eq:A in D gauge}\end{equation}
We expect the discretized Lagrangian in this gauge to produce a stable
algorithm, at $\alpha\neq1/2$, as long as the particle remains near
to $\left(x,y\right)=\left(0,0\right)$. %
This is illustrated in Figure~\ref{fig:D gauge example}, where equations
of motion are integrated explicitly ($\alpha=0$) for differing
initial conditions. The nonlinear motion close to $\left(0,0\right)$
is stable, while with initial conditions further from $\left(0,0\right)$
the integrator blows up. We emphasize that the algorithm is stable
for any initial condition at $\alpha=1/2$ and that the purpose of this example
is to show the gauge change can be used locally to give a stable \emph{explicit}
integrator. Of course, more complicated particle trajectories would
preclude the use of such a linearization technique: particles would
quickly move into regions where a different discretization gauge was
necessary. Future investigations could include exploring the possibility of
stitching together local gauges to give a globally stable, nonlinear, explicit
algorithm.

\section{Electromagnetic gauge\label{sec:Electromagnetic-gauge}}

In any physical system related to electromagnetism, dynamics must
be invariant under an electromagnetic gauge transformation, $\bm{A}\left(\bm{x},t\right)\rightarrow\bm{A}\left(\bm{x},t\right)+\nabla\lambda\left(\bm{x},t\right)$,
$\phi\left(\bm{x},t\right)\rightarrow\phi\left(\bm{x},t\right)-\partial_{t}\lambda\left(\bm{x},t\right)$.
For the case of the single particle guiding center Lagrangian, Eq.~\eqref{eq:Full GC Lagrangian},
such a transformation is of course a special case of the gauge transformations
considered in the previous section. Evidently continuous
particle dynamics are invariant under a change of electromagnetic
gauge. However, we have just illustrated that stability properties of the
variationally discretized system can be strongly altered by gauge changes.
Unlike traditional algorithms, in which the equations of motion (and
thus $\bm{B}\left(\bm{x}\right)$ and $\bm{E}\left(\bm{x}\right)$)
are discretized, the variational symplectic guiding center integrators
are\emph{ not} electromagnetically gauge invariant. This can be illustrated 
explicitly (for the algorithm of Refs.~\onlinecite{Qin:2008p5812,Qin:2009p5777})
by making the transformation $A_{i}^{\dagger}\rightarrow A_{i}^{\dagger}+\lambda_{,\, i}$, 
$\phi\rightarrow \phi-\partial_t\lambda$ 
in Eq.~\eqref{eq:VSI algorithm 1}. This leads to the extra term,
\begin{equation}
\frac{1}{2h}\left[\lambda_{,\, ij}\left(x_{k+1}^{i}-x_{k-1}^{i}\right)-\left(\lambda_{,\, j}\left(\bm{x}_{k+1}\right)-\lambda_{,\, j}\left(\bm{x}_{k-1}\right)\right)\right]+\partial_t \lambda_{,\,j}\left(\bm{x}_k\right),\label{eq:gauge trans of VSI eqns}
\end{equation}
which is non-zero (but does of course vanish in the continuous limit).
It is important to explore this further to understand limitations of the algorithm 
and how best to choose a gauge to obtain a reasonable approximation
of particle motion.

The preceding considerations provide compelling motivation to: (i) 
restore gauge invariance to the discrete Lagrangian, (ii)
compare this gauge invariant algorithm to the integrators from Refs.~\onlinecite{Qin:2009p5777,Qin:2008p5812,Li:2011p7068},
and (iii), determine the conditions under which they should give a valid description of the motion. This
can be achieved by replacing evaluations of $\bm{A}$ and $\phi$
at a single spacetime point (for instance $\left(\bm{x}_{k}+\bm{x}_{k+1}\right)/2$)
with time integrals over a particle trajectory. For example, a discretized
version of Eq.~\eqref{eq:Full GC Lagrangian} that is gauge invariant
is \begin{align}
L_{d} & =\left[\int_{t_{k}}^{t_{k+1}}\frac{dt}{h}\bm{A}\left[\bm{x}\left(t\right)\right]+U_{k+\nicefrac{1}{2}}\bm{b}\left(\bm{x}_{k+\nicefrac{1}{2}}\right)\right].\left(\bm{x}_{k+1}-\bm{x}_{k}\right)/h\nonumber \\
 & -\left[\int_{t_{k}}^{t_{k+1}}\frac{dt}{h}\phi\left[\bm{x}\left(t\right)\right]+\Gamma\left(\bm{x}_{k+\nicefrac{1}{2}},U_{k+\nicefrac{1}{2}}\right)\right].\label{eq:Gauge invar discrete L}\end{align}
Here, $\bm{x}_{k+\nicefrac{1}{2}}$ indicates $\left(\bm{x}_{k}+\bm{x}_{k+1}\right)/2$
and the path in the time integral, $\bm{x}\left(t\right)$, is
simply a straight line between $\bm{x}_{k}$ and $\bm{x}_{k+1}$,
that is, $\bm{x}\left(t\right)=\bm{x}_{k}+\left(\bm{x}_{k+1}-\bm{x}_{k}\right)\left(t-t_{k}\right)/h$.
To prove gauge invariance of discrete equations of motion, we need
to show that the discrete action, $\mathcal{S}_{d}=\sum_{k}L_{d}$,
is unchanged (except at the endpoints) by an electromagnetic gauge
transformation. For Eq.~\eqref{eq:Gauge invar discrete L}, 
first note that $\left(\bm{x}_{k+1}-\bm{x}_{k}\right)/h$ is $\bm{v}\left(t\right)$.
The gauge transformation thus amounts to the addition of \begin{equation}
\int_{t_{k}}^{t_{k+1}}\frac{dt}{h}\bm{v}.\nabla\lambda\left(\bm{x}\left(t\right),t\right)+\int_{t_{k}}^{t_{k+1}}\frac{dt}{h}\frac{\partial\lambda\left(\bm{x}\left(t\right),t\right)}{\partial t}\label{eq:Gauge trans in D Lag}\end{equation}
to Eq.~\eqref{eq:Gauge invar discrete L}. The first term is 
\begin{equation}
\int_{t_{k}}^{t_{k+1}}\frac{dt}{h}\left[\frac{d\lambda}{dt}-\frac{\partial\lambda}{\partial t}\right],\label{eq:Left over term}
\end{equation}
the second part of which cancels the second term of Eq.~\eqref{eq:Gauge trans in D Lag}.
Carrying out the integral, we are left with \begin{equation}
\lambda\left(\bm{x}_{k+1},t_{k+1}\right)-\lambda\left(\bm{x}_{k},t_{k}\right),\label{eq:Terms left from gauge trans}\end{equation}
which contributes $\lambda\left(\bm{x}_{N},t_{N}\right)-\lambda\left(\bm{x}_{0},t_{0}\right)$
$ $to $\mathcal{S}_{d}$. Since this is only a boundary contribution,
the discrete equations of motion are unchanged and thus electromagnetically
gauge invariant. Note that in a numerical implementation of the algorithm
obtained from Eq.~\eqref{eq:Gauge invar discrete L}, the time integrals
would need to be evaluated numerically. This calculation could be
exact for piecewise polynomial $\bm{A}$ and $\phi$ (using Gaussian
quadrature), as would be the case if they were defined discretely
on some grid. Such discrete fields are used in many applications and an 
electromagnetically gauge invariant algorithm as introduced here could  easily  be 
implemented. As a side note, this is particularly important for use in a variational particle-in-cell 
scheme, where a particle pusher is coupled to an electromagnetic field solver in 
a single discrete variational principle. Ensuring electromagnetic gauge invariance 
of the discrete Lagrangian guarantees that the scheme satisfies the current continuity equation, 
$\partial_t \rho+\nabla \cdot \bm{J}=0$, which implies Gauss's law remains satisfied  
at all times\cite{PRLtobePublished}.

\subsection{Numerical example}

The variational guiding center integrators considered use the discretizations
$\bm{A}\left(\bm{x}\right)\rightarrow \frac{1}{2}\left[\bm{A}\left(\bm{x}_{k}\right)+\bm{A}\left(\bm{x}_{k+1}\right)\right]$ 
(Refs.~\onlinecite{Qin:2008p5812,Qin:2009p5777}) or $\bm{A}\left(\bm{x}\right)\rightarrow\bm{A}\left(\bm{x}_{k+\nicefrac{1}{2}}\right)$ 
(Ref.~\onlinecite{Li:2011p7068}). We would expect the lack of gauge invariance
to be relatively unimportant if these terms (and similar terms for
$\phi$) were good approximations to $\int_{t_{k}}^{t_{k+1}}\frac{dt}{h}\bm{A}\left[\bm{x}\left(t\right)\right]$,
which is essentially an average of $\bm{A}$ over the particle
trajectory. Thus, to minimise the consequences of the lack of electromagnetic
gauge invariance on numerical results, we should choose a gauge such
that the resulting $\bm{A}$ and $\phi$ are as smooth as possible.
We note that this idea gives an answer to the question of how to implement
the variational guiding center algorithms for a given magnetic field,
perhaps defined on a grid. To ensure a stable algorithm, one should
choose an $\bm{A}\left(\bm{x}\right)$ that is as smooth
as possible under the constraint $\nabla\times\bm{A}=\bm{B}$.

\begin{figure}
\begin{centering}
\includegraphics{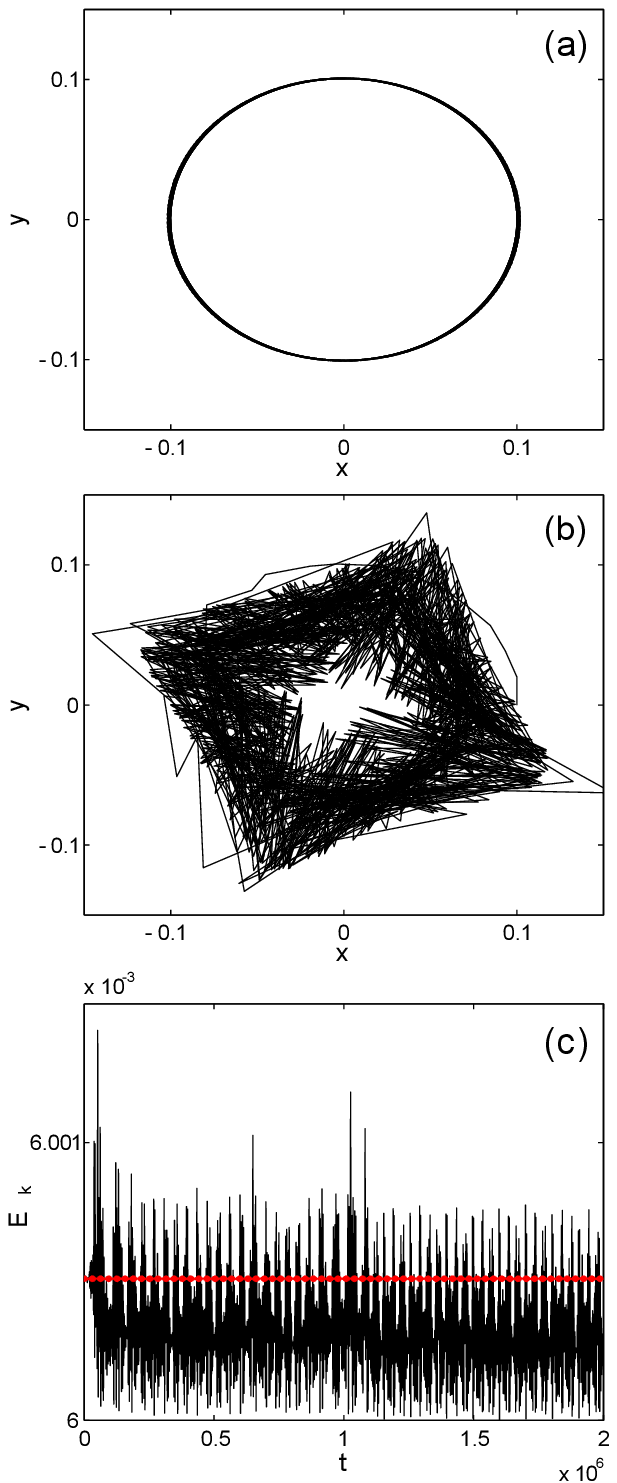}
\caption{Numerically integrated particle trajectories in the field $\bm{B}\left(\bm{x}\right)=\left[1+\left(x^{2}+y^{2}\right)/20\right]\bm{\hat{z}}$:
a) Using $\bm{A}=\left(-\frac{1}{60}y^{3},\, x+\frac{1}{60}x^{3},\,0\right)$,
b) using $\bm{A}=\left(-\frac{1}{60}y^{3},\, x+\frac{1}{60}x^{3},\,0\right)+\nabla\cos\left(10\, xy\right)$.
The kinetic energy, $E_{k}\left(t\right)$ is plotted in c), with
the trajectory of a) in red, and that of b) in black. The time-step $h$ is chosen
so that the particle rotates by approximately $1/10$ radians per
timestep and the trajectory is integrated for 1000 timesteps.\label{fig:EM gauge change}}
\end{centering}

\end{figure}

We test out this idea numerically in Figure~\ref{fig:EM gauge change}.
This shows integrated guiding center particle trajectories for same
magnetic field as the previous example, $ $$\bm{B}\left(\bm{x}\right)=\left[1+\left(x^{2}+y^{2}\right)/20\right]\bm{\hat{z}}$.
As before $\bm{A}=\left(-\frac{1}{60}y^{3},\, x+\frac{1}{60}x^{3},\,0\right)$
is used in Figure~\ref{fig:EM gauge change}(a), while in Figure~\ref{fig:EM gauge change}(b)
we gauge transform this  $\bm{A}$ with $\lambda=\cos\left(10\,x\,y\right)$.
For the parameters of Figure~\ref{fig:EM gauge change}(b), there
will be a relatively large change in $\bm{A}\left(\bm{x}\right)$
over a timestep, meaning $\bm{A}\left(\bm{x}_{k+\nicefrac{1}{2}}\right)$
will not necessarily be an accurate approximation to $\int_{t_{k}}^{t_{k+1}}\frac{dt}{h}\bm{A}\left[\bm{x}\left(t\right)\right]$.
This manifests itself in a highly unstable particle trajectory and
kinetic energy [Figure~\ref{fig:EM gauge change}(c)]. This property
of the variational guiding center algorithms should not be 
problematic in practice provided a relatively smooth gauge is chosen
and the time-step is sufficiently small. Numerical investigations have revealed
that, as long as  $\bm{A}\left(\bm{x}\right)$
and $\phi\left(\bm{x}\right)$ are not unusually uneven, timestep restictions are less
severe than for conventional algorithms, such as  fourth order
Runga-Kutta.

\section{Conclusions and future work}

The linear stability properties of the variational symplectic
guiding center algorithms in Refs.~\onlinecite{Qin:2008p5812,Qin:2009p5777,Li:2011p7068}
have been systematically examined to provide new insights into 
how these relate to gauge transformations of the governing Lagrangian.
It was found that an oddity in the relationship between the discrete
and continuous symplectic forms explains why explicit variational 
guiding center integrators have been observed to be numerically unstable. This can be mitigated
by the use of an antisymmetric discretization gauge, in which even an explicit
integrator is stable. However, this gauge does not always exist globally for
realistic fields in general co-ordinates. Results from investigation of the
consequences of the lack of electromagnetic gauge invariance in the
variational symplectic guiding center algorithm  indicate that as long as $\bm{A}\left(\bm{x}\right)$ is relatively
smooth, the algorithm is approximately gauge invariant and should accurately 
reproduce particle dynamics.

There are still numerous properties and instabilities of the variational
guiding center algorithm that require future work. One such instability,
referred to in Ref.~\onlinecite{Li:2011p7068}, affects the integrated parallel
velocity, $u$, for fully 3-dimensional fields. The velocity is seen
to oscillate between even and odd time-steps, with the amplitude growing
in time. This instability is nonlinear, a complication for a systematic
analysis, but can be mitigated by formulating the algorithm in terms
of $u_{k+1/2}\equiv\left(u_{k}+u_{k+1}\right)/2$ rather than $u_{k}$.
Another area of ongoing research is in variational integrators for
fields defined discretely on a grid, as would be required, for example,
if the magnetic field is output from another code. Preliminary results
show certain numerical instabilities associated with the piecewise nature of
$\bm{A}$. The results presented above on electromagnetic gauge 
transformations may be important in such studies, and investigations into 
gauge invariant integrators are ongoing. 

\section*{Acknowledgements}

This research is supported by U.S.~DOE (DE-AC02-09CH11466).


%

\end{document}